\documentclass[prb,aps,amsmath,amssymb,floatfix,superscriptaddress,longbibliography]{revtex4-2}
\usepackage{blindtext}
\usepackage{subcaption}
\usepackage{caption}
\usepackage{float}
\usepackage{soul,color}
\usepackage{graphicx}
\usepackage{bm}
\newcommand{\bea}{\begin{eqnarray}}
\newcommand{\eea}{\end{eqnarray}}
\usepackage[colorlinks=true, citecolor=blue, linkcolor=blue, urlcolor=blue]{hyperref}
\usepackage[utf8]{inputenc}
\usepackage[T1]{fontenc}
\usepackage{mathptmx}
\usepackage{caption}
\usepackage{subcaption}
\newcommand{\ninej}[9]{
\ensuremath{
\left\{\!\!
\begin{array}{ccc}
#1 & #2 & #3 \\
#4 & #5 & #6 \\
#7 & #8 & #9 \\
\end{array}
\!\!\right\}
}}

\begin{document}

\title[]{ Cold collisions between alkali and alkaline-earth hetero nuclear atom-ion system Li + Ba$^+$}
\author{Dibyendu Sardar} \email{chem.dibyandu.sardar@gmail.com}
\affiliation{JILA, University of Colorado, Boulder, Colorado 80309, USA}
\author{Somnath Naskar}
\affiliation{Department of Physics, Jogesh Chandra Chaudhuri College, Kolkata-700033, India}
\date{\today}
\begin{abstract}

In a recent experiment by Schaetz group \cite{weckesser2021}, the quantum $s$-wave regime has been attained for alkali and alkaline-earth atom-ion combination (Li-Ba$^+$). We investigate possible outcomes from interaction of this ion-atom pair at quantum regimes from a theoretical point of view. For this purpose,  Born-Oppenheimer potential energy surfaces are constructed for the lowest three dissociation channels of (Ba-Li)$^+$ molecular system using a multireference configuration interaction (MRCI) electronic structure calculation. We present elastic, spin-exchange (SE), and diffusion cross sections at different energy regimes. The collisional properties of this system are calculated in terms of the scattering phase shifts and scattering cross sections, and the semiclassical behavior at a relatively large energy limit is also examined. For SE collisions, phase locking is obtained towards lower partial waves.
\end{abstract}

\maketitle

\section{Introduction}
Synthesization of cold molecules or molecular ions from cold atoms or ion-atom (IA) mixture is one of the major thirst among the research topics in the domain of atomic and molecular physics. Amidst several methods available for the formation of cold molecules, two of the most important methods are Photoassociation (PA) \cite{weiner} and Magnetoassociation (MA) \cite{kohler2006} at ultracold temperatures. The rich structure of the cold molecular ions purveys many new applications and research directions from precision measurements to quantum computing and quantum simulation \cite{carr}. In the realm of ultracold temperatures, where the de Broglie wavelength becomes comparable to or longer than the particle size or inter-particle separation, such systems exhibit several quantum effects like resonances, tunneling, etc. This opens up a new door for understanding controlled chemical reactions at ultracold energies. 

The IA combination has received considerable attention and impressive evolution over the last decade. By harnessing the mutual interaction between the two quantum systems, namely atom and ion, an integrated IA hybrid setup has been formed \cite{schmid2010}. The interaction between an ionic species and a neutral particle is governed by the electrical induction process. It can be understood in terms of interaction of charge of an ion with the electrons of neutral atoms. Generally, these induction-controlled interactions are stronger compared to van der Waals types of interactions. Whenever a neutral atom comes near an ion, the atom is polarized by the electric field of ion i.e. the ion induces dipole moment in the atom and thereby interacts with it. The IA interaction potential is given by $V(R) = -C_4/R^4$, where $C_4$ is the induction coefficient and depends on the static polarizability of the atom. These hybrid systems may offer a new platform for investigating elastic, inelastic, and reactive collisions between ions and atoms at low temperatures. IA collisions are important to understand charge transport phenomena \cite{cote2000classical}, IA bound states \cite{PhysRevLett.89.093001}, cold-molecular ions \cite{sardar2016} etc. Recently, SE reaction processes has been investigated in IA colliding systems \cite{ratschbacher2012,ratschbacher2013decoherence}. It has been proposed that controlled IA cold collisions may be used for future quantum information processing \cite{doerk2010atom}. 

Over the last couple of years, there have been several studies on alkali atom and alkaline-earth ion systems. The frequent use of alkaline-earth ionic species in most of the hybrid IA experiments is due to its suitability for laser cooling which aids to achieve low IA collision energies. Some of the important and well studied heteronuclear alkali-alkaline earth IA systems are Na-Be$^+$ \cite{ladjimi2018}, Na-Ca$^+$ \cite{makarov2003}, Rb-Ca$^+$ \cite{belyaev2012}, Rb-Yb$^+$ \cite{mclaughlin2014}, Li-Yb$^+$ \cite{tomza2015cold}, K-Mg$^+$ \cite{farjallah2019}, Cs-Mg$^+$ \cite{farjallah2022electronic}. Apart from heteronuclear IA combination, studies have also been made on homonuclear alkaline-alkaline earth systems, eg, Be-Be$^+$ \cite{zhang2011charge}, Mg-Mg$^+$ \cite{alharzali2018}, and Yb-Yb$^+$ \cite{zhang2009scattering}. In most of these studies, cold collisions, formation of the molecular ions and Feshbach resonances (FR) are the major goals accomplished either by PA or via MA. But, experimental realization of such important phenomena is yet to achieve. 

A usual experimental obligation of IA systems towards achieving low temperatures is that the ions cannot be cooled to $\mu$K or sub$\mu$K temperatures due to the presence of inherent trap-induced micro-motion. Heavier the ion is, lesser is the micro-motion and thereby more suitable for cooling. Theoretically, it has been proposed that the lowest energy regimes may be reached for a IA combination having the highest mass ratio \cite{cetina2012micromotion}. In a very recent experiment, the quantum regime has been reached with an alkali-alkaline earth IA system having mass ratio $\sim 23-28$. The system Li-Yb$^+$ \cite{furst2018dynamics,feldker2020buffer} is an early workhorse in this effort. In this system $s$-wave regime is attained at a collision energy of 8.6$k_B$ $\mu$K, where $k_B$ is Boltzmann constant. The investigation comprises spin-dynamics of single trapped Yb$^+$ ion in a cold spin-polarized bath of Li atoms without any signature of FR in such energy limit. 

In the experiment by Schaetz group \cite{weckesser2021}, FR have been detected in case of a single trapped $^{138}$Ba$^+$ ion and $^6$Li atoms. For this system, a total number of eleven FR have been identified out of which four are due to $s$-wave FR for different values of tunable magnetic field. These observed number of resonances is mainly due to the additional interaction namely the second order spin-orbit coupling (SOC). This  coupling mixes internal states ($m_F$) with the rotational motion ($l, m_l$), causing the increased number of resonances \cite{ticknor2004} in $^6$Li-$^{138}$Ba$^+$ system. These encouraging results provide a deeper insight into IA interactions, paving thereby a way to explore complex many body systems and quantum simulations. For low energy domains, especially in $\mu$K or sub-$\mu$K regime, quantum mechanical scattering calculations are inevitable to describe the IA interactions, where the scattering is characterised by quantum phase shift in scattered wave function. Quantities like scattering length at the s-wave regime, scattering cross section etc can be calculated in terms of this quantum phase shift. At a relatively larger energy limit, $E\ge1 $mK$\times k_B$, a simpler semiclassical description is useful.

In most of the alkali-alkaline earth IA systems the choice of initial collision channel lies at the excited asymptote with lighter-atom-heavier-ion combination. This choice is a matter of experimental compulsion as heavier ion is favourable for cooling as discussed earlier. A disadvantage of this situation is that the excited asymptote is short-lived and decays to ground asymptotic channel through radiative charge exchange collisions. As an interesting exception, however, the collisional asymptote of Li-Ba$^+$ is energetically lower than that of Li$^+$-Ba. Thus, the initial collision channel for (BaLi)$^+$ IA system is naturally chosen to be the Li-Ba$^+$ ground collisional asymptote \cite{weckesser2021} which is free from radiative charge transfer loss mechanism and therefore having much much longer lifetime as compared to other IA systems. Along with this, as discussed earlier, this particular IA system allows an ultracold window to study controllable collision resonances. Thus, from experimental point of view, compared to other IA systems, the $^6$Li-$^{138}$Ba$^+$ system turns out to be extremely novel one. 

Depending on the orientation of the electronic spin located at each species in their ground state, two molecular potentials are formed - a singlet-sigma (X$^1$$\Sigma^+$) and a triplet-sigma (a$^3$$\Sigma^+$) at the short range IA internuclear separation. These two potentials offer elastic collisions and inelastic SE collsions at low energy regimes. In the presence of second order SOC, the total spin projection is not conserved during collision tendering the possibility of spin relaxation (SR) \cite{sikorsky2018spin}. The SR competes with SE process and weakens the spin-control of the system. In the ultracold regime, however, the rate of SR becomes much slower \cite{sikorsky2018phase} than the Langevin collision rate and the former is strongly suppressed. While studying SE collisions at low energies, partial wave phase-locking (PWPL) \cite{sikorsky2018phase} has been observed where difference in the quantum phase-shifts due to the two potentials found to be independent of partial wave quantum number. 

In this paper, our aim of theoretical investigation with the $^6$Li-$^{138}$Ba$^+$ system is the following. In Section \ref{sec-ab}, we exploit the \textit{ab initio} method for calculating electronic structures of the (BaLi)$^+$ system. We present in Section \ref{sec-FR}, a detailed prospective scheme  to realise FR in the system. Section \ref{sec-collision} is devoted in investigating IA collisions employing the electronic states X$^1$$\Sigma^+$ and $a^3\Sigma^+$ which go asymptotically to the same dissociation limit.  We determine the collisional properties in terms of scattering phase shift and scattering cross section and attempt to justify the semiclassical behaviour at a relatively large energy limit. We also study the PWPL effect at low energies. We conclude in Section \ref{sec-conclude}.

\section{\textit{ab initio} calculation}
\label{sec-ab}
\begin{table}
\centering
\caption{First ionization potential values of Li and Ba in EV. }
\begin{tabular}{ |c|c|c|c| } 

\hline
  Atom & IP (EV) & Expt & Theory\\
\hline
Li & 5.393 & 5.4$ \pm 0.2$ \cite{yu2006nist} &   5.39 \cite{yu2006nist} \\ 
Ba & 5.057 & 5.3 $\pm 0.3$ \cite{yu2006nist}   & 5.0 \cite{yu2006nist} \\ 
\hline
\end{tabular}
\label{tb1}
\end{table}
In this section, we describe \textit{ab initio} calculations to construct the potential energy surface of a heteronuclear IA system (BaLi)$^+$. The \textit{ab initio} calculations are performed by MOLPRO 2012.1 software package \cite{werner2008}. The electronic structure calculations are performed by the multireference configuration interaction (MRCI) with additional Davidson correction that approximately accounts for the size consistency and higher excitations. 

The ground state electronic configurations of the atoms Ba and Li are expressed as: [Kr]$_{36}$4d$^{10}$5s$^2$5p$^6$6s$^2$ and 1s$^2$2s$^1$, respectively.  The lighter candidate lithium is described by the correlation consistent polarized valence quadruple-$\zeta$ basis set with augmenting functions e.g., aug-cc-pwCVQZ \cite{prascher2011}. For the Barium atom, a pseudopotential-based correlation consistent polarized weighted core valence triple-$\zeta$ basis set (cc-pwCVTZ-PP) is used \cite{hill2017}, where the inner core electrons are described by the Stuttgart / Koeln effective core potential ([ECP46MDF] \cite{lim2005}).

The inner core electrons of Ba are replaced by core potentials ECP46MDF, leaving eight sub-valance and two valence electrons in the outer shell. Therefore, an effective number of molecular orbitals of (BaLi)$^+$ is reduced to seven where twelve electrons are distributed. These occupied molecular orbitals are expressed as (5$a_1$, 1$b_1$, 1$b_2$, 0$a_2$) or (5, 1, 1, 0) in the C$_{2v}$ Abelian point group symmetry used by MOLPRO. We define an active space denoted as (9, 4, 4, 1) in which the 5d6p atomic orbitals of Ba and 2p atomic orbitals of Li are included in the reference space. The doubly occupied orbitals are set to (3, 1, 1, 0). The energies of the molecular orbitals on this active space are calculated in the following way: Initially, the spin-restricted optimized Hartree-Fock (HF) molecular orbitals are used as solutions for the complete active space self-consistent field (CASSCF) problem. Thereafter, a dynamical correlation is carried out by internally contracted MRCI with single and double excitations being taken relative to this CASSCF reference wave functions where the 5s5p atomic orbitals of Ba and 1s orbital of Li are correlated. The above described method is used in constructing final potential energy surfaces, nevertheless we further verify the validity of approach. 

It is a difficult task to estimate the uncertainty of \textit{ab initio} calculations, especially for a many-electrons system including a heavy atom. However, before proceeding to describe the potential energy curves of the (BaLi)${^+}$ molecular system, we compare the results of our computed atomic components to the available experimental data. The comparison is accomplished in terms of the first ionization potential (IP) of Ba and Li, including static electric dipole polarizability ($\alpha$) for both neutral and charged components of Barium and lithium atoms as shown in Table.\ref{tb1} and \ref{tb2}. In our current method of calculation, the predicted ionization potential of Li and Ba agree well with previously reported experimental and theoretical value with an error of less than 1\%. With regards to the static electric dipole polarizability of the ground state Li, our calculated value is in good agreement with the experimental result \cite{molof1974} but differs by 0.3 $a_0^3$ from the theoretical value \cite{tomza2015cold}. The calculated value of $\alpha$ for Li$^+$ is 0.192 $a_0^3$ and the experimental value is 0.188$ \pm 0.002$ \cite{cooke1977}. As for the Ba and Ba$^+$ components, the  static polarizabilities are in good agreement with experimental and theoretical values. The harmony of these results with literature values can impart a reliable description of the diatom molecular ion (BaLi)$^+$ using our current level of theory and the basis sets.

\begin{figure}
  \includegraphics[width=0.8\linewidth]{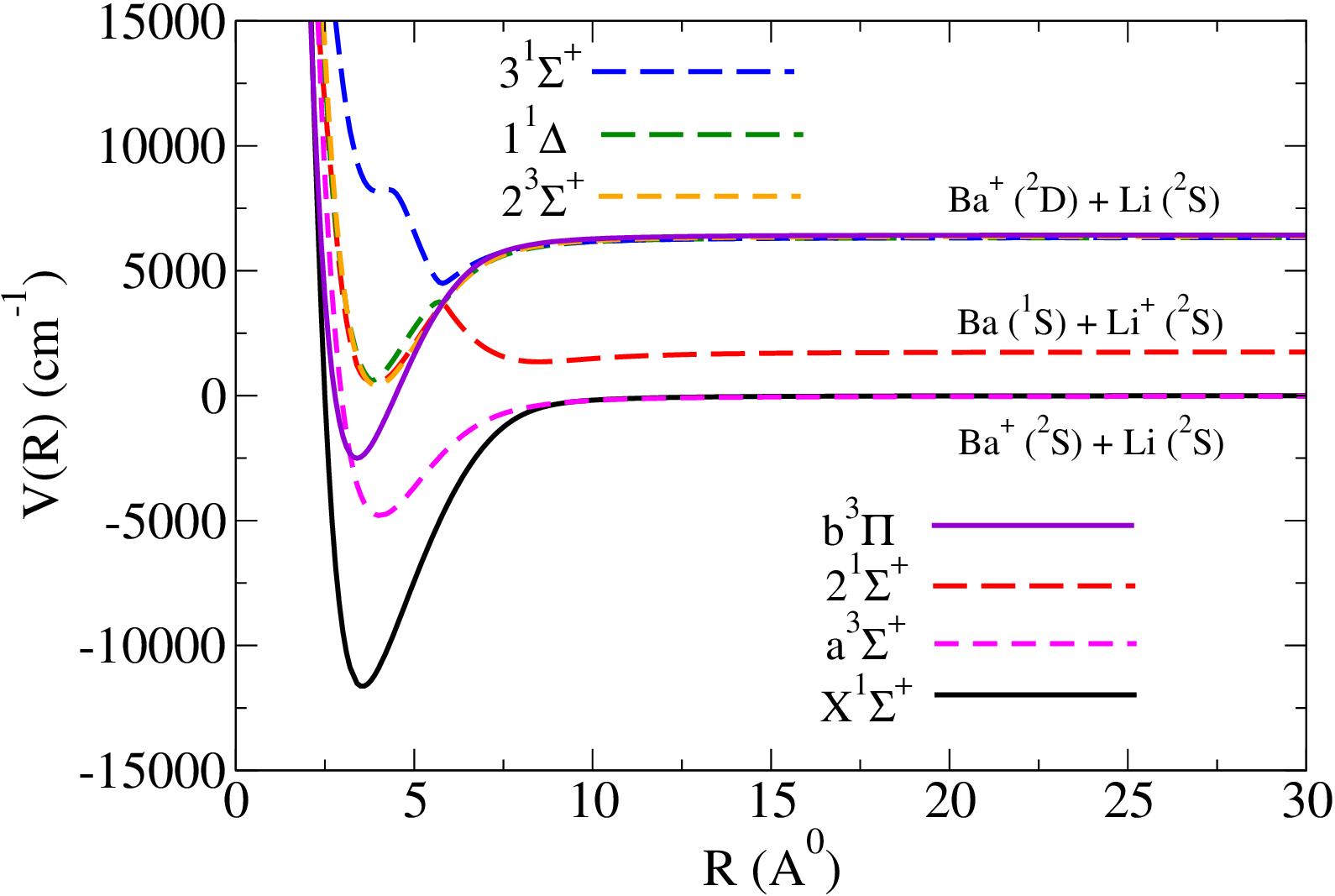}
  \caption{The adiabatic potential energy curves of lowest three dissociation channels of (BaLi)$^+$ molecular system are plotted as a function of IA internuclear separation. The energies are in wave number.}
  \label{fig1}
\end{figure}

\begin{table}
\centering
\caption{Static electric dipole polarizability values of Li (Li$^+$) and Ba (Ba$^+$) in atomic unit. }
\begin{tabular}{ |c|c|c|c| } 

\hline
  Atom/ion & $\alpha$ (a.u)& Expt & Theory\\
\hline
Li & 164.30 & 164 $\pm 3.4$ \cite{molof1974}&   164.0 \cite{tomza2015cold}\\ 
 Li$^+$ & 0.192 & 0.188$ \pm 0.002$ \cite{cooke1977}&  0.190 \cite{tomza2015cold} \\ 
Ba & 267.74 & 268 $\pm 6$ \cite{miller1978atomic}    & 268.19 \cite{sahoo2008}\\ 
Ba$^+$ & 124.22 & 123.88 $\pm 5$ \cite{snow2007} &  124.26 $\pm 1$  \cite{sahoo2009}\\ 
\hline
\end{tabular}
\label{tb2}
\end{table}
Some of the potential energy curves for the ground and excited states of (BaLi)${^+}$ molecular system are shown in FIG. \ref{fig1}. The spectroscopic parameters associated with these potentials are represented in Table.\ref{tb3} in terms of the equilibrium bond length (R$_e$) and depth of the well ($D_e$). All the equilibrium bond lengths of these molecular electronic states are expressed in Angstrom whereas the $D_e$'s are in cm$^{-1}$.
For this alkali and alkaline-earth IA system, depending on the location of the positive charge at the dissociation limit, two possible IA combination may arise - either alkali-ion-alkaline-atom or alkali-atom-alkaline-ion. The corresponding dissociation threshold will be energetically different depending on the ionization potential of the monomers involved. In the latter case, there are two unpaired valence electrons in each species that results in singlet and triplet molecular potential curves.

For the case of (BaLi)${^+}$, the possible asymptotic arrangements are given as Ba$^+$-Li and Ba-Li$^+$ depending on the localization of the positive charge either the Ba atom or Li atom. Since the first ionization potential of Ba is lower than Li as compared in Table.\ref{tb1}, making Ba$^+$-Li arrangement is in the absolute ground state asymptote of (BaLi)${^+}$ system. Due to this feature of the ground state asymptote, this system is significantly different compared to other heteronuclear alkali alkaline-earth IA systems, as mentioned earlier.   
\begin{table}[h!]
\centering
\caption{The spectroscopic constants for some lowest molecular electronic states of BaLi$^{+}$ are expressed in terms of equilibrium bond length and depth of the well. }
\begin{tabular}{ |c|c|c|c| } 

\hline
 Molecular state & R$_\text{e}$ (A$^0$) & D$_\text{e}$ (cm$^{-1}$)\\
\hline
X$^1$$\Sigma^+$ & 3.55 & 11627 \\ 
a$^3$$\Sigma^+$ & 4.00 & 4784  \\ 
2$^1$$\Sigma^+$ & 4.01 & 1206 \\ 
3$^1$$\Sigma^+$ & 5.80 & 1833 \\ 
2$^3$$\Sigma^+$ & 3.96 & 5961  \\ 
b$^3$$\Pi$ & 3.40 & 8935  \\ 
1$^1$$\Delta$ & 3.80 & 5729  \\ 
\hline
\end{tabular}
\label{tb3}
\end{table}
In the ground state asymptotic arrangement of the (BaLi)${^+}$ system, the interaction occurs between the ground state Ba$^+$ ion and ground state Li atom which are both open shells. Two electronic states 
X$^1$$\Sigma^+$ and a$^3$$\Sigma^+$ appear as a result of this interaction. The features of these two states need to be explained as we use them frequently in the domain of cold collisions.  
The singlet-sigma state is strongly bound having binding energy equal to 11627 cm$^{-1}$ and the equilibrium position is located at 3.58 A$^{0}$. On the other hand, the triplet-sigma state is bound by 4675 cm$^{-1}$, and the equilibrium distance is equal to 4 A$^{0}$. 

The two other dissociating thresholds conceive the electronic states 2$^1$$\Sigma^+$, 3$^1$$\Sigma^+$, 2$^3$$\Sigma^+$, b$^3$$\Pi$ and 1$^1$$\Delta$. 
The asymptotes Ba ($^1$S) + Li$^+$($^2$S) and Ba$^+$($^2$D) + Li($^2$S) lie above the ground state asymptote by an energy $\sim$1740 cm$^{-1}$ and $\sim$6300 cm$^{-1}$, respectively. There is one noticeable pattern in FIG. \ref{fig1}. The electronic state a$^3$$\Sigma^+$ correlated to the ground state asymptote crosses the state b$^3$$\Pi$ at a location near 3.2 A$^0$. This curve crossing facilitates a large spin-orbit coupling which plays an important role in the spin non-conserving scattering phenomena. In the following we mention the method and result of spin-orbit matrix element calculation.

\subsection*{Spin-orbit matrix element}
\begin{figure}
  \includegraphics[width=0.8\linewidth]{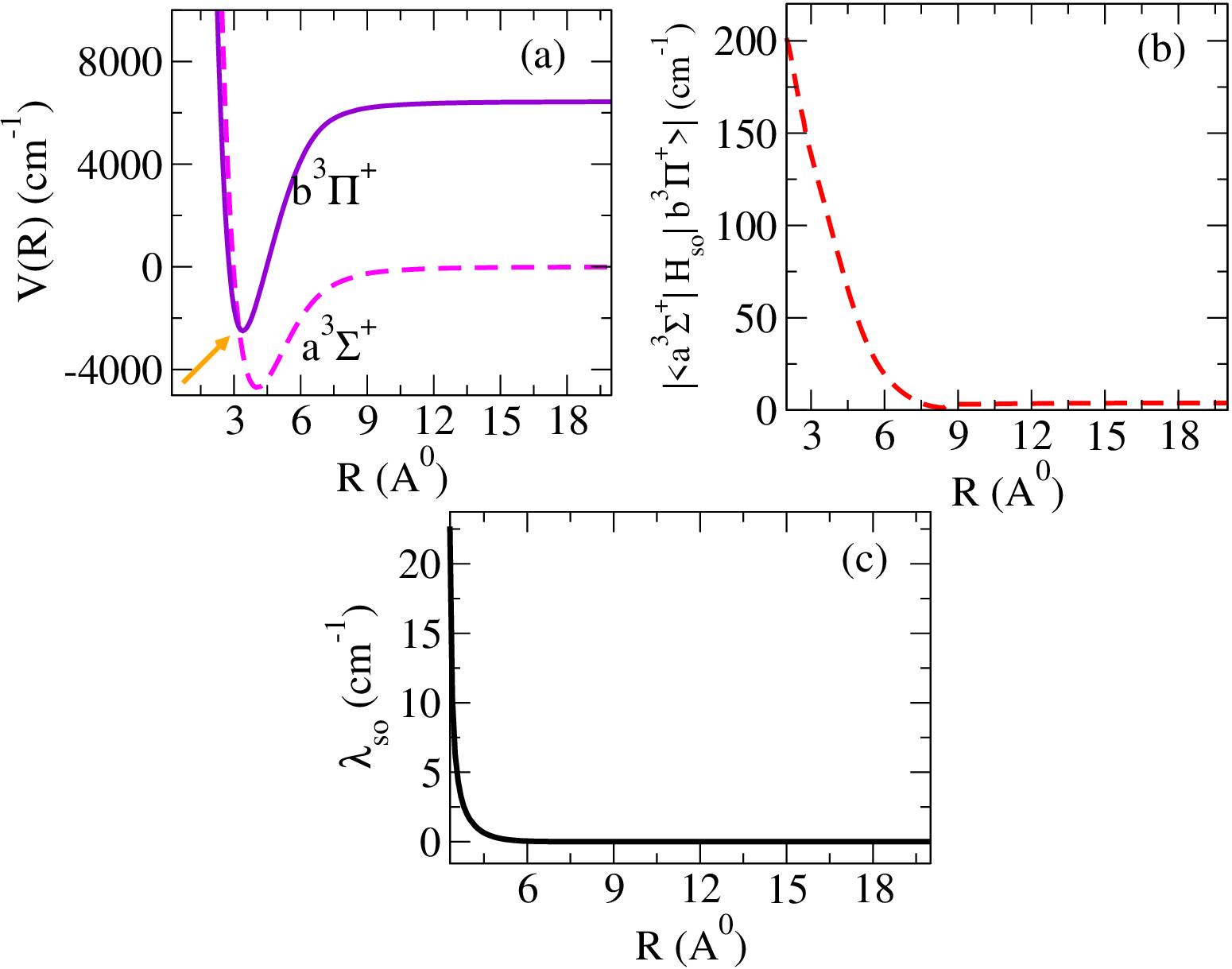}
  \caption{The potentials a$^3$$\Sigma^+$ and b$^3$$\Pi$ are presented in panel-(a) and the arrow indicates the crossing between these two potentials. The spin-orbit matrix element and spin-orbit coupling constant are shown in panel-(b) and (c), respectively.} 
  \label{fig2}
\end{figure}
 The spin-orbit coupling matrix elements ($\xi^{SO}$) between two molecular states a$^3$$\Sigma^+$ and b$^3$$\Pi$ can be expressed as
\begin{equation}
\xi^{SO} = \langle a^3\Sigma^+|H_{SO}|b^3\Pi \rangle    
\end{equation}
where $H_{SO}$ is the spin-orbit Hamiltonian which is either the Breit-Pauli (BP) operator or spin-orbit pseudopotentials (ECPs). The matrix elements of the spin-orbit coupling Hamiltonian are evaluated by exploiting the wave functions of electronic states a$^3$$\Sigma^+$ and b$^3$$\Pi$. The calculations of the electronic wave functions are carried out by MOLPRO. The spin-orbit coupling constant ($\lambda$) associated with the spin-orbit matrix element is
\begin{equation}
    \lambda (R) = \frac{2}{3}\frac{|\langle a^3\Sigma^+|H_{SO}|b^3\Pi \rangle|^2}{V_{b^3\Pi}(R) - a^3\Sigma^+ (R)}
\end{equation}
It is noteworthy to mention that the sign of the originally calculated spin-orbit matrix elements is not well defined as the phase of the corresponding wave functions is arbitrary. It meant that the phase of the coupling matrix elements strictly depends on the signs of the
successive calculations performed: it starts with the relative signs of the molecular orbitals
optimized by CASSCF, which are used in the MRCI calculations on the triplet a$^3$$\Sigma^+$ and b$^3$$\Pi$ states, providing MRCI eigenvectors each defined with a phase factor. The spin-orbit integrals are calculated using the CASSCF MOs (with their signs) and the spin-orbit matrix elements
combine the CI eigenvectors with the integrals. Each of the calculations performed on a given geometry is thus correct, but the relative signs between different geometries are arbitrary. 

In panel-(b) of FIG. \ref{fig2} we plot the absolute value of the spin-orbit matrix element in cm$^{-1}$ as a function of internuclear separation between the electronic states a$^3$$\Sigma^+$ and b$^3$$\Pi$. We note that the matrix element decreases exponentially with increase in IA distance. In panel-(a) of FIG. \ref{fig2} we plot the two concerned potentials as a function of IA distance showing the crossing between the curves. The curve crossing occurs at a distance R = 3.2 A$^{0}$. Finally, the  calculated second-order spin-orbit coupling constant is shown in the panel-(c) which decays exponentially with IA distance. We note that the value of $\lambda_{SO}$ near the equilibrium distance of a$^3$$\Sigma^+$ potential is comparable to (LiYb)$^+$ system. 

\section{Construction of hyperfine potentials: prospective of Feshbach resonances}
\label{sec-FR}
Experimentally the ion Ba$^+$ is prepared in an incoherent mixture spin state $\left|6S_{1/2};\hspace{1mm} s^{Ba^+} = 1/2,\hspace{1mm} m_s^{Ba^+} = \pm {1/2}\right \rangle$ and the atom Li is considered in hyperfine state $\left|f_{Li} = 1,\hspace{1mm} m_f^{Li} = - 1/2\right\rangle$. Then the ion and the atom are allowed to interact, resulting a number of FR out of which four are detected as $s$-wave resonances. 
In this section we introduce a brief description of the construction of hyperfine potentials that could be used to characterise the $s$-wave FR.
\subsection*{Channel classification}
The ion $^{138}$Ba$^+$ has vanishing nuclear spin ($i_1 = 0$) with electronic spin $s_1 = 1/2$.
One can label the hyperfine sub levels as $\left|f_1 = 1/2,\hspace{1mm} m_{f_1} = \pm 1/2\right\rangle$. On the other hand $^6$Li has nuclear spin $i_2 = 1$ with hyperfine quantum number $f_2 = 3/2$ and 1/2;  and these two hyperfine levels are separated by an energy 228.2 MHz ($\sim$ 10.952 mK). In the presence of an external magnetic field, the projection $M_J$ of total angular momentum ${\bf J} = {\bf f_1} + {\bf f_2} + {\bf l}$ remains a good quantum number during a collision event. In addition, states with different orbital angular momentum (${\bf l}$) characterizing different partial waves are decoupled if one neglects the anisotropic spin-spin interaction, making $l$ and its projection $m_l$ conserved. Under this condition, $M_f = m_{f_1} + m_{f_2}$ will also be conserved. Here, we restrict our discussion in the subspace of partial wave $l = 0$ including $M_f$ as constant. We constrict the collision within the sub block $M_f = 1/2$ as it represents the lowest energy channel state. In the absence of magnetic field B=0, For $M_f = 1/2$, there are four possible channels constituting the asymptotic or uncoupled basis $\left|f_1 m_1, f_2 m_2\right\rangle$ as listed in Table.\ref{tb4}.
\begin{figure}
  \includegraphics[width=0.6\linewidth]{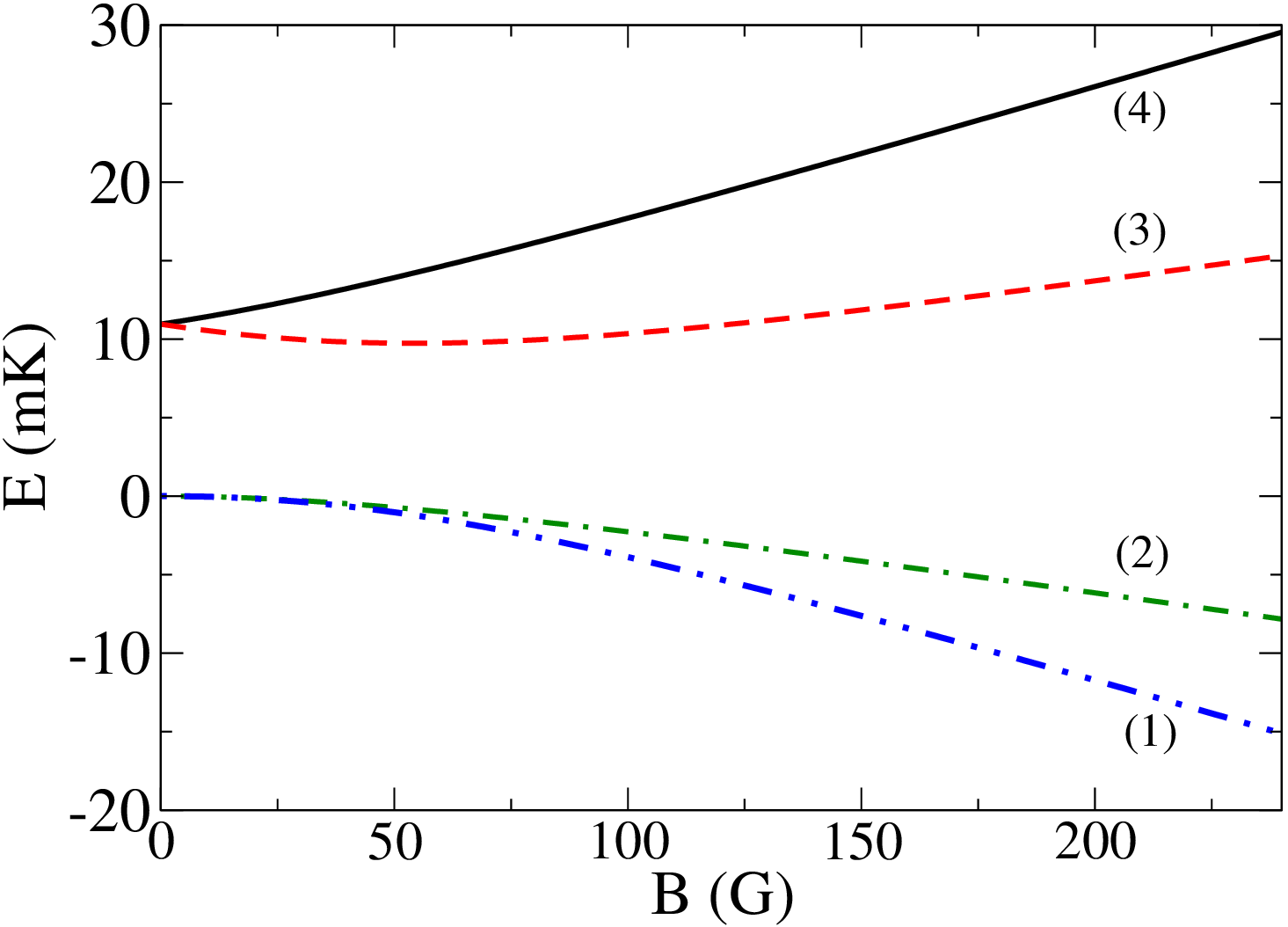}
  \caption{The variation of energy of the four channels as a function of magnetic field.}
  \label{fig3}
\end{figure}

\begin{figure}
  \includegraphics[width=0.7\linewidth]{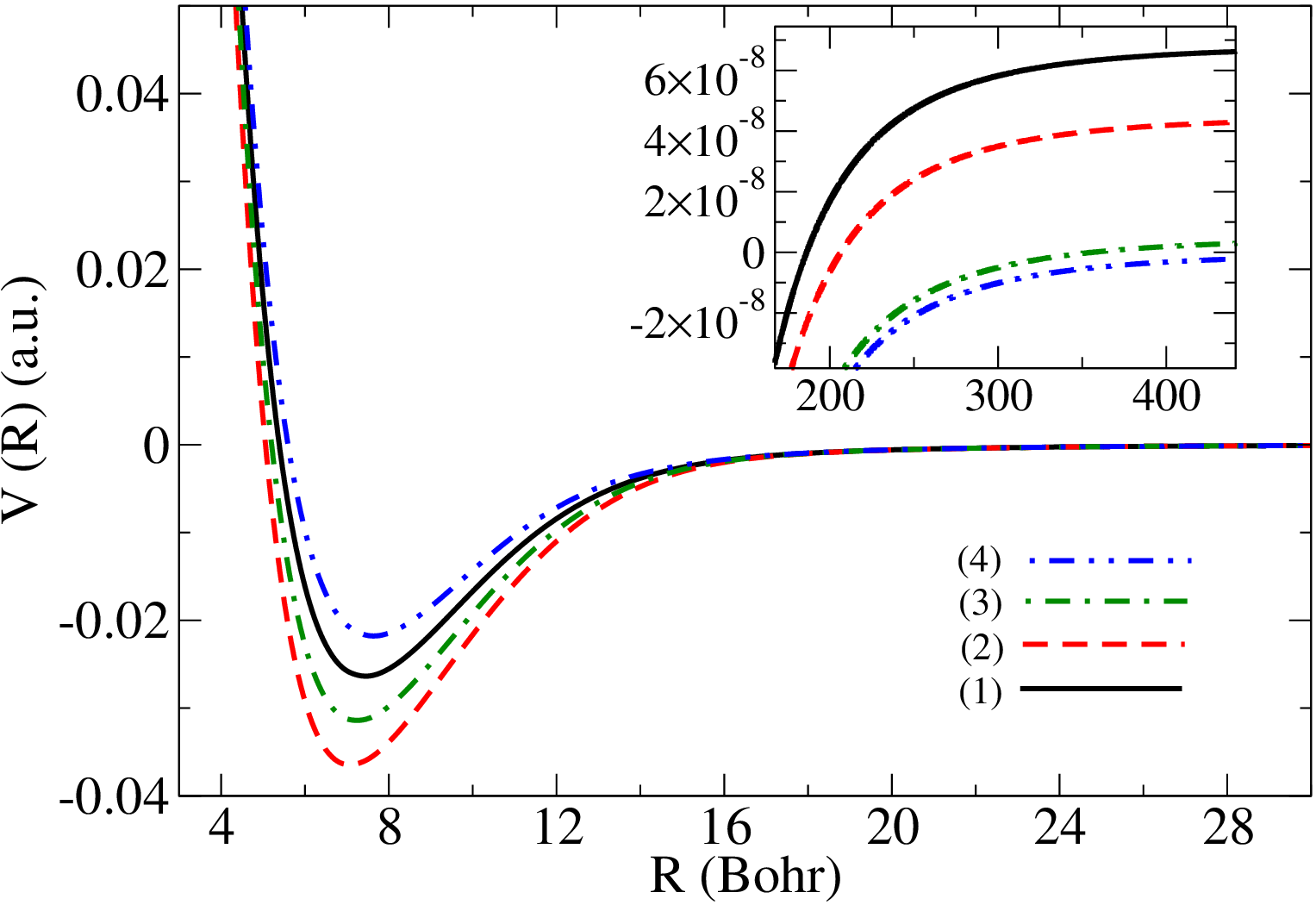}
  \caption{The variation of four diagonal potentials of (BaLi)$^+$ system as a function of IA distance at short range for a particular magnetic filed B = 100 G. The corresponding asymptotes are shown in the inset of the figure. }
  \label{fig4}
\end{figure}

In the absence of magnetic field, the hyperfine interaction is diagonal in the basis $ \left|f_1 m_1, f_2 m_2\right\rangle$. In this condition, since both $f^2$ and $l^2$ are conserved, another useful basis that may be used is the coupled hyperfine basis $ \left|(f_1 f_2)f m_f\right\rangle$. One can use this basis in the presence of a weak magnetic field considering a perturbative treatment with a very small Zeeman interaction. In short notation, we denote this basis as $ \mid b\rangle$. Now in the presence of magnetic field, let the basis that diagonalizes both Zeeman and hyperfine terms be $\mid\tilde{b}\rangle$. On diagonalization, one obtains eigen values that define the threshold energies of the channels and the eigen vectors that are related to the basis $\mid b\rangle$ through some linear transformation as
\begin{equation}
    \mid b\rangle = \sum_{\tilde{b}} \langle\tilde{b} \mid b \rangle \mid \tilde{b} \rangle 
\end{equation}
The matrix elements of the central potential in the coupled basis can be given as 
\begin{align}
\left\langle (f_1 f_2) f m_f \mid V^c \mid (f_1' f_2') f' m_f' \right\rangle = 
\sum_{S,I,M_S,M_I} V_S  \left\langle (f_1 f_2) f m_f \mid S M_S, I M_I  \right\rangle \left\langle S M_S, I M_I \mid (f_1' f_2') f' m_f' \right\rangle
\end{align}
Here $\mid S M_S, I M_I  \rangle$ is the adiabatic basis and the central potential is diagonal in this basis with eigenvalues $V_S$. The central potential can be written as: $V^c = V_0(r)P_0 + V_1(r)P_1$; where $V_0(r)$ and $V_1(r)$ correspond to the singlet X$^1$$\Sigma^+$ and triplet a$^3$$\Sigma^+$ states, respectively. $P_0$ and $P_1$ are the corresponding projections.
\begin{table}[h!]
\centering
\caption{Four asymptotic channels for $M_F = 1/2$ of ($^{138}$Ba-$^{6}$Li)$^+$ system. }
\begin{tabular}{|c| c| c| c| c|c| } 
 \hline
 channels & $(f_1, m_{f_1})$ & $(f_2, m_{f_2})$& E$^{\infty}$ (mK)  \\ [0.5ex] 
 \hline
 1 & $(1/2, 1/2)$ & $(1/2, -1/2)$ & 0  \\ 
 2 & $(1/2, -1/2)$ & $(1/2, 1/2)$ & 0 \\
 3 & $(1/2, 1/2)$ & $(3/2, -1/2)$ & 10.952 \\
 4 & $(1/2, -1/2)$ & $(3/2, 1/2)$ & 10.952 \\[1ex]   
 \hline
\end{tabular}
\label{tb4}
\end{table}
Here  the adiabatic basis and the coupled asymptotic basis are related through the transformation matrix elements
\begin{equation}
\left\langle S M_S, I M_I \mid (f_1 f_2) f m_f \right\rangle =  C^{S,I,f}_{M_S,M_I,m_f}
\sqrt{(2 f_1 + 1)(2 f_2 + 1)(2 S + 1)(2 I + 1)}
\ninej{s_1}{i_1}{f_1}{s_2}{i_2}{f_2}{S}{I}{f}
\left( \frac{1+(1-\delta_{f_1 f_2})(-1)^{S+I+l}}{\sqrt{2-\delta_{f_1 f_2}}}\right)
\end{equation}
where $C^{S,I,f}_{M_S,M_I,m_f}$ is the Clebsch-Gordan (CG) coefficient and $m_f = M_S + M_I$. The quantity in the curly bracket is known as  $9j$-symbol \cite{edmonds1996angular}. 
Considering these transformations, we present the variation of channel energies with magnetic field in FIG. \ref{fig3} where the numbers inside the plot indicate the indices of the channels. In FIG. \ref{fig4} we present four diagonal potentials in short range regime as a function of internuclear separation $R$ (Bohr) for a particular magnetic field B = 100 G. The asymptotic long range part of the potentials are shown in the inset of FIG. \ref{fig4} and the energy of the said channels increases from channel-(4) to (1) as a function of magnetic field. With these primary knowledge at hand, one can study the FR in (Ba-Li)$^+$ system.
\section{Cold collisions}
\label{sec-collision}
\begin{figure}
     \centering
     \begin{subfigure}[b]{0.49\textwidth}
         \centering
         \includegraphics[width=\textwidth]{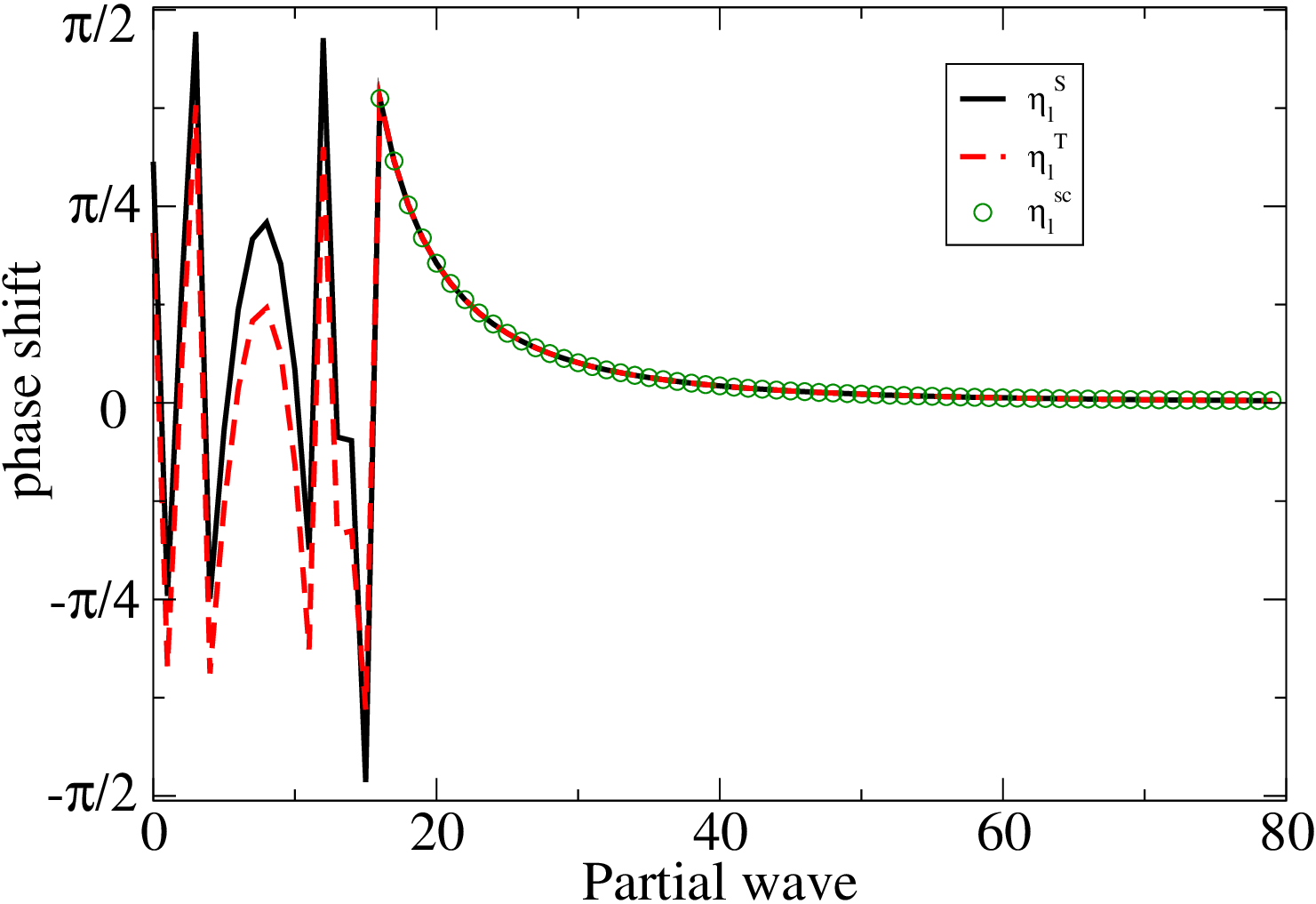}
         \label{fig:y equals x}
     \end{subfigure}
     \hfill
     \begin{subfigure}[b]{0.49\textwidth}
         \centering
         \includegraphics[width=\textwidth]{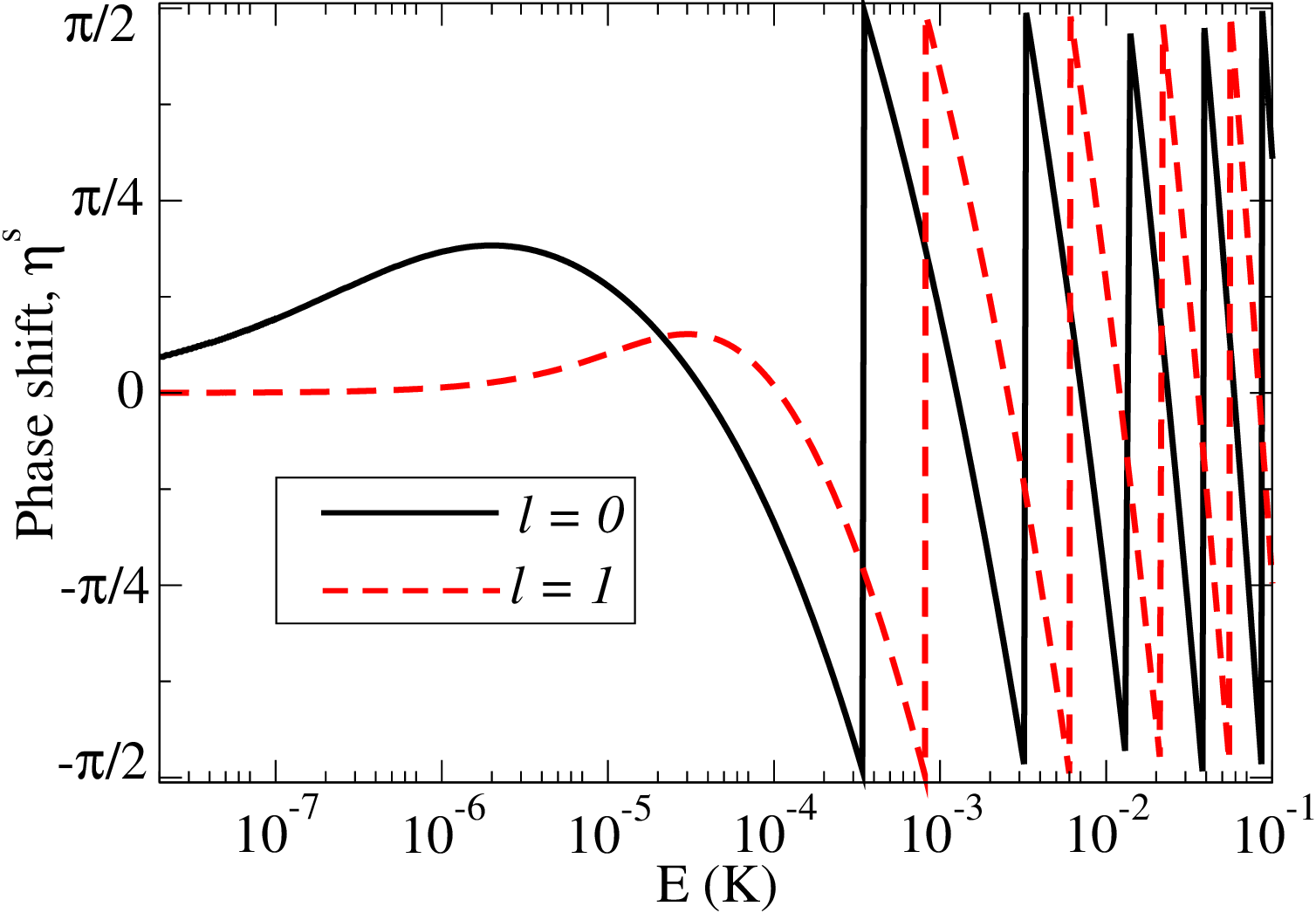}
         \label{fig:three sin x}
     \end{subfigure}
     \caption{In the left-hand panel quantum and semiclassical phase shifts are plotted as a function of partial waves for the potentials X$^1$$\Sigma^+$ and a$^3$$\Sigma^+$, respectively. The quantum phase shifts are shown for the partial waves $l = 0, 1$ as a function of energy only for X$^1$$\Sigma^+$ state in right-hand panel.}
        \label{phase}
\end{figure}
Experimentally the ion $^{138}$Ba$^+$ and atom $^6$Li both are prepared in their electronic ground state $^2$S which corresponds to the lowest energy dissociation channel (Ba$^+$ ($^2$S) + Li ($^2$S)) of (BaLi)$^+$ system. Therefore the charge exchange collision requiring a photon or enough collision energy corresponding to an excitation of the system for the next dissociation limit are strongly suppressed. Herein, we consider mainly low energy processes in the domain of energy $\mu$K to sub-mK and therefore we restrict our discussions only to the area of elastic scattering and SE processes. 

\subsection{Elastic Collision}
\begin{figure}
  \includegraphics[width=0.7\linewidth]{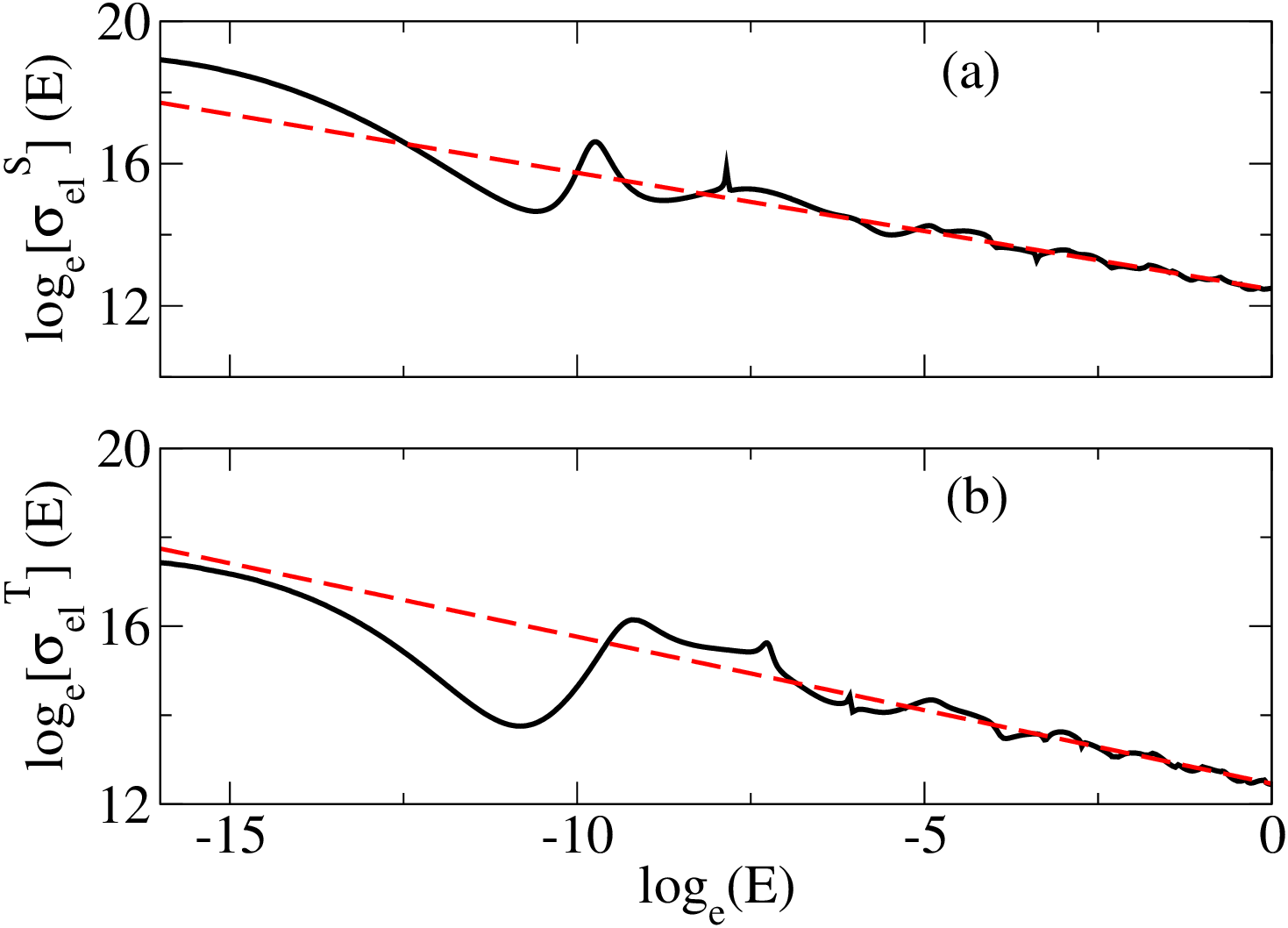}
  \caption{Total elastic scattering cross sections are plotted as a function of collision energy in the log-log scale for the concerned singlet and triplet potentials in the panel-(a) and (b), respectively. The dashed line indicates the semiclassical fitting in cross sections.}
  \label{elastic}
\end{figure}
An ion $^{138}$Ba$^+$ collides  elastically with a neutral atom $^6$Li in their ground state is associated with the asymptote (Ba$^+$ ($^2$S) + Li ($^2$S)) where both the potentials X$^1$$\Sigma^+$ and a$^3$$\Sigma^+$ will be relevant. Applying the method of partial wave decomposition in the total wave function, the time independent Schr\"{o}dinger equation at a given collision energy E is given by
\begin{equation}
    \left[\frac{d^2}{dR^2} + k^2 -2\mu V_{S,T} (R) - \frac{l(l+1)}{R^2}\right] y_{E,l}^{S,T} (R)=0
    \label{eq1}
\end{equation}
where wave number $k=\sqrt{2\mu E}/\hbar$, $\mu$ is the reduced mass of the colliding IA pair, and $l$ is the partial wave. The quantity $V_{S}$ and $V_{T}$ are related with the potential X$^1$$\Sigma^+$ and a$^3$$\Sigma^+$, respectively. The long-range part of the potential is approximated as $V_{S,T}(R) = -\left(\frac{C_4}{R^4} + \frac{C_6}{R^6}\right)$; where $C_4 = \frac{1}{2}q^2\alpha_{Li} $, the coefficient related to the dipole-polarizability ($\alpha_{Li}$) of Li atom. The quadrupole polarizability ($\beta_{Li}$) is associated with $C_6$ coefficient, $C_6 = \frac{1}{2}q^2\beta_{Li}$. In this work we use $\alpha_{Li} = 164.1$ a.u and $\beta_{Li} = 1424$ a.u. The short range and long range part of the potentials are joined smoothly by cubic spline algorithm. 
In the asymptotic limit, the wave function can be expressed in terms of Bassel ($j_l$(kr)) and Neumann ($n_l$(kr)) functions as
\begin{equation}
y_{E,l}^{S,T} (R) = kR\left[j_l(kr)\cos\eta_l^{S,T}-n_l(kr)\sin\eta_l^{S,T}\right]
\end{equation}
where $\eta_l^{S}$ and  $\eta_l^{T}$ are the phase shifts associated with the potentials X$^1$$\Sigma^+$ and a$^3$$\Sigma^+$, respectively. The equation.\ref{eq1} is solved numerically by Numerov-Cooley method using three-point recursion relation. The details of this method is discussed in Ref.\cite{alharzali2018}. 

In the left hand panel of FIG. \ref{phase}, we present the plot of phase shifts by varying the number of partial waves for the potentials X$^1$$\Sigma^+$ and a$^3$$\Sigma^+$, considering the collision energy E = 0.1 K. For a given large collision energy and high partial waves, the potentials $V_{S,T} (R)$ behave, to the leading term as $-C_4/R^4$. Under this condition, one can find the semi-classical phase shift  $\eta_l^{SC} \simeq (\pi \mu^2\alpha_{Li})/(4\hbar^4)\times E/l^3$. In our concerned potentials, the semiclassical phase shift is in agreement with the quantum phase shifts for partial waves $l > 21$ as shown in FIG. \ref{phase} with collision energy E = 0.1K. In the right hand panel of FIG. \ref{phase}, we show quantum phase shifts for first two partial waves ($l = 0, 1$) as a function of collision energy for the PES X$^1$$\Sigma^+$. Note that the $s$-wave ($l = 0$) phase shift is dominant at very low energies where that of $p$-wave ceases to zero. The phase shifts for both $s$ and $p$-wave change sign, indicating the presence of pole where scattering length diverges.

The scattering phase shift is associated with another physically measurable quantity; the scattering cross section. For a direct elastic collision, the cross sections ($\sigma_{el}^{S,T}$) can be expressed in terms of scattering amplitude as 
\begin{equation}
    \sigma_{el}^{S,T} (E) = \int \mid f_{S,T} \mid^2 d\Omega = \frac{4\pi}{k^2}\sum_{l=0}^{\infty}(2l+1)\sin^2\left(\eta_l^{S,T}\right)
\end{equation}
where $\mid f_{S,T} \mid^2$ are the scattering amplitudes associated with the singlet and triplet potentials of (BaLi)$^+$ system and $d\Omega$ is the differential solid angle. At a relatively large collision energy, the cross sections are approximated by the semi-classical expression 
\begin{equation}
    \sigma_{sc} (E) = \pi \left(\mu\alpha_{Li}^2/\hbar^2\right)^{1/3} \left(1 + \pi^2/16\right) E^{-1/3}.  
    \label{eq8}
\end{equation}
Thus the plot of $\left[\log{\sigma_{sc}(E)}\text{ vs }\log E\right]$ is a straight line with slope -1/3 and the intercept is associated with dipole polarizability of neutral atom Li. 

We present total elastic scattering cross sections as a function of energy in Kelvin for the X${^1}\Sigma^{+}$ and a$^3\Sigma^+$ potentials in the panel-(a) and (b) of FIG. \ref{elastic}. The cross section includes the sum of the first 81 partial waves for the concerned singlet sigma state and 65 partial waves for the triplet sigma state. Here we note that the $s$-wave contribution is dominant in all the cases at energies corresponding to temperatures 0.1 $\mu$K. For neutral alkali atom systems, however, the $s$-wave contribution is dominant for energies around 100 $\mu$K. This is due to the existence of the long range polarization potential in the IA system as compared to the shorter range van der Waals interactions between neutral atoms. 
As energy increases, more and more partial waves start to contribute to the total elastic scattering cross sections. In order to check the convergence for the sum over partial waves, we fit both the plots linearly at the large energy limit. The numerically calculated slope is quite close to the semi-classical theoretical value, confirming the convergence of partial wave contributions in the total elastic scattering cross section.
\begin{figure}
     \centering
     \begin{subfigure}[H]{0.49\textwidth}
         \centering
         \includegraphics[width=\textwidth]{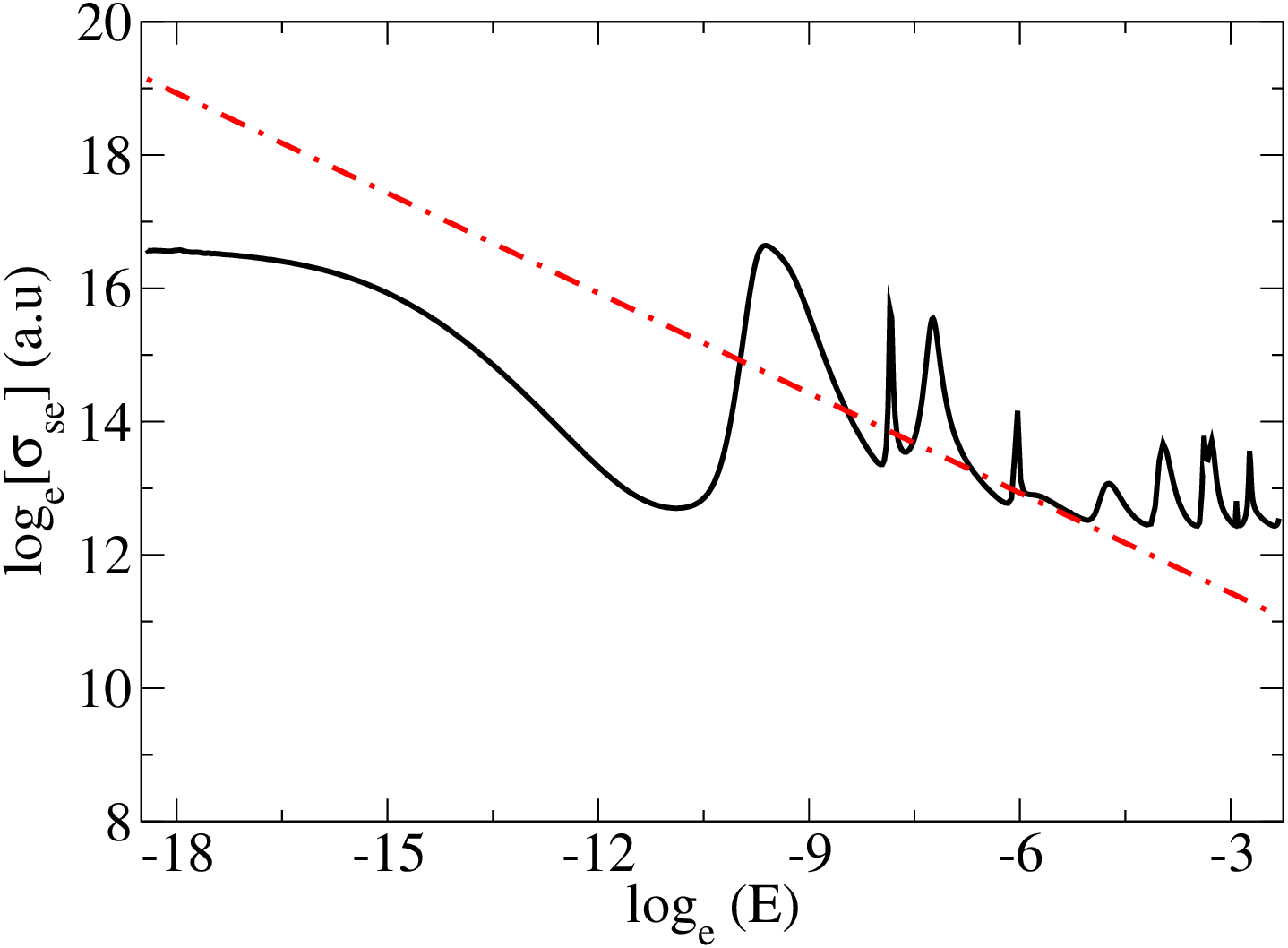}
         \caption{SE cross section as a function of energy in the log-log scale with.}
         \label{se}
     \end{subfigure}
     \hfill
     \begin{subfigure}[H]{0.49\textwidth}
         \centering
         \includegraphics[width=\textwidth]{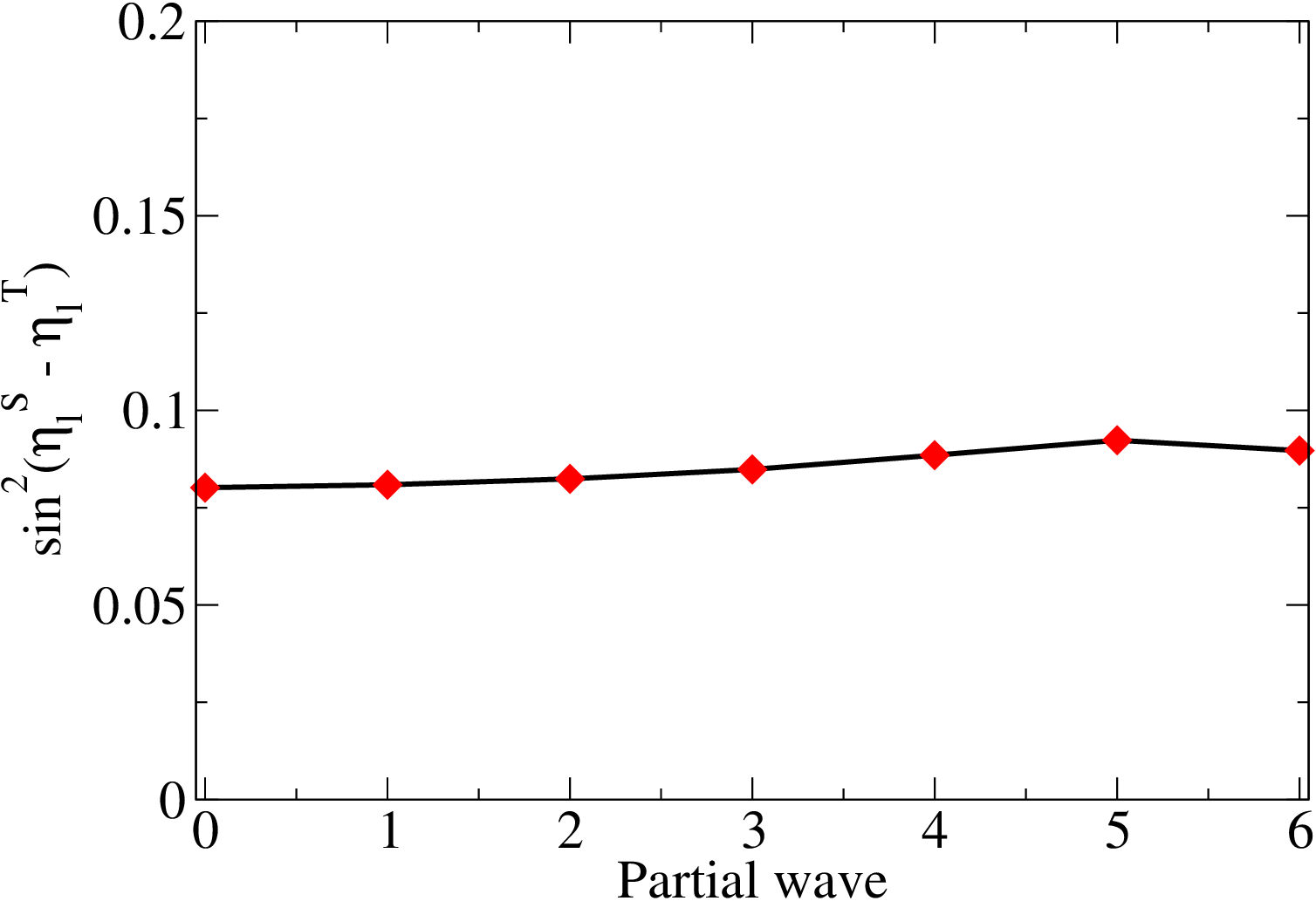}
         \caption{The difference in phase shift in radian
as a function of \textit{l} at energy $E$ = 10 mK.}
         \label{fig:pwpl}
     \end{subfigure}
     \caption{Study of spin-dynamics }
        \label{spin-spin}
\end{figure}

\subsection{Spin Exchange (SE) Collision}
At the short range, another possible outcome of the collisions between the ultra cold ion $^{138}$Ba$^+$ and atom $^6$Li in their ground state is SE collision. Under SE interaction, the total spin projection is conserved along any axis. Due to this conservation property, if the ion and the atom are prepared with parallel electronic spins, they can interact only on the triplet potential and no SE takes place. On the other hand, if they are prepared with anti-parallel spins, they interact on both singlet and triplet potentials opening up a finite probability of SE. In the elastic as well as the degenerate internal state approximation, this scattering event is described in terms of the singlet and triplet scattering phase shifts. In the short range, the inelastic SE cross section can be expressed as \cite{makarov2003}
\begin{equation}
    \sigma_{SE} (E) = \frac{\pi}{k^2}\sum_{l=0}^{\infty}(l+1)\sin^2\left(\eta_l^S-\eta_l^T\right)
\end{equation}
In the Langevin regime of intermediate collisional energy, the SE cross section shows the classical Langevin behaviour: $\sigma_{L} (E) \propto E ^{-1/2}$. In FIG. \ref{se} we show SE cross section as a function of collision energy in kelvin in logarithm scale. The curve can approximately be fitted with a Langevin behaviour (red dashed line) with slope -1/2 in the energy range (0.56 - 7.01) mK. 

Due to the short-range nature of the SE interaction, the phase difference $\left(\eta_l^S-\eta_l^T\right)$ remains constant for over a range of $l$ values exhibiting PWPL \cite{sikorsky2018phase}. For the (Ba-Li)$^+$ system, at 10 mK of collisional energy, the maximum partial wave number ($l_{max}$) that contributes to $\sigma_{SE} (E)$ is found to be 7. We calculate $\sin^2\left(\eta_l^S-\eta_l^T\right)$ for these partial waves which remains constant ($\cal{C}$) with 5.5\% standard deviation as shown in FIG. \ref{fig:pwpl}. Thus under PWPL approximation, the expression for SE cross section simplifies to 
\begin{equation}
\sigma_{SE} (E)=\frac{3\pi \cal{C}}{2k^2}l_{max}\left(l_{max}+1\right)  
\end{equation}

Conservation of the total spin projection is violated in the presence of SOC as it is not diagonal in the adiabatic basis $\left|\hspace{.8mm} S M_S, I M_I  \hspace{.8mm}\right\rangle$. This leads to SR. Thus the spin dynamics is governed by the competition of SE and SR. When the SR rate is significant compared to the Langevin collision rate, the spin-controllability of the system is lost. However, In the ultracold regime, where the steady state temperature is much lower than the hyperfine energy gap (10.952 mK), the SR rate becomes significantly small.
\begin{figure}
     \centering
     \begin{subfigure}[H]{0.49\textwidth}
         \centering
         \includegraphics[width=\textwidth]{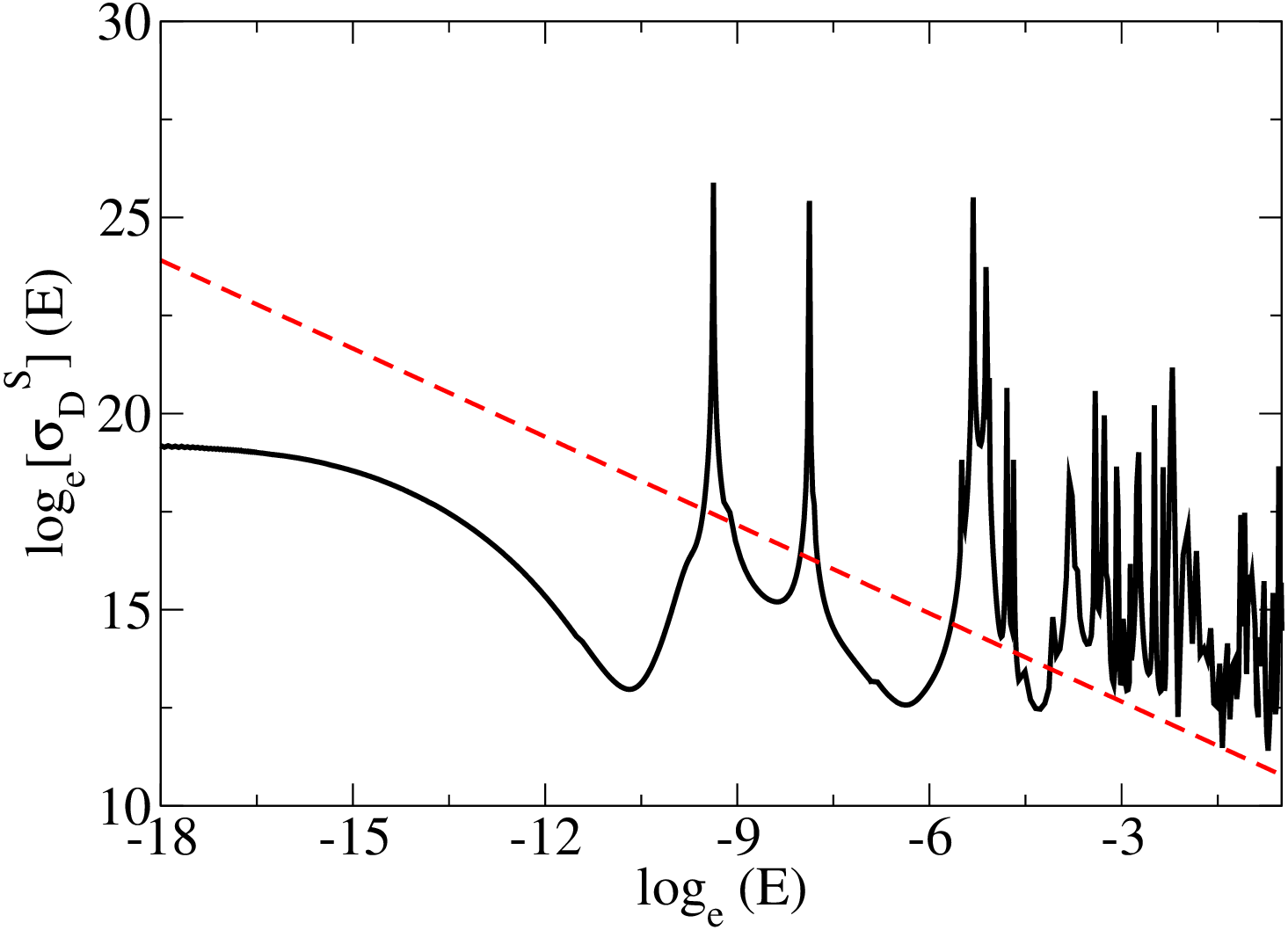}
         \label{fig:y equals xn}
     \end{subfigure}
     \hfill
     \begin{subfigure}[H]{0.49\textwidth}
         \centering
         \includegraphics[width=\textwidth]{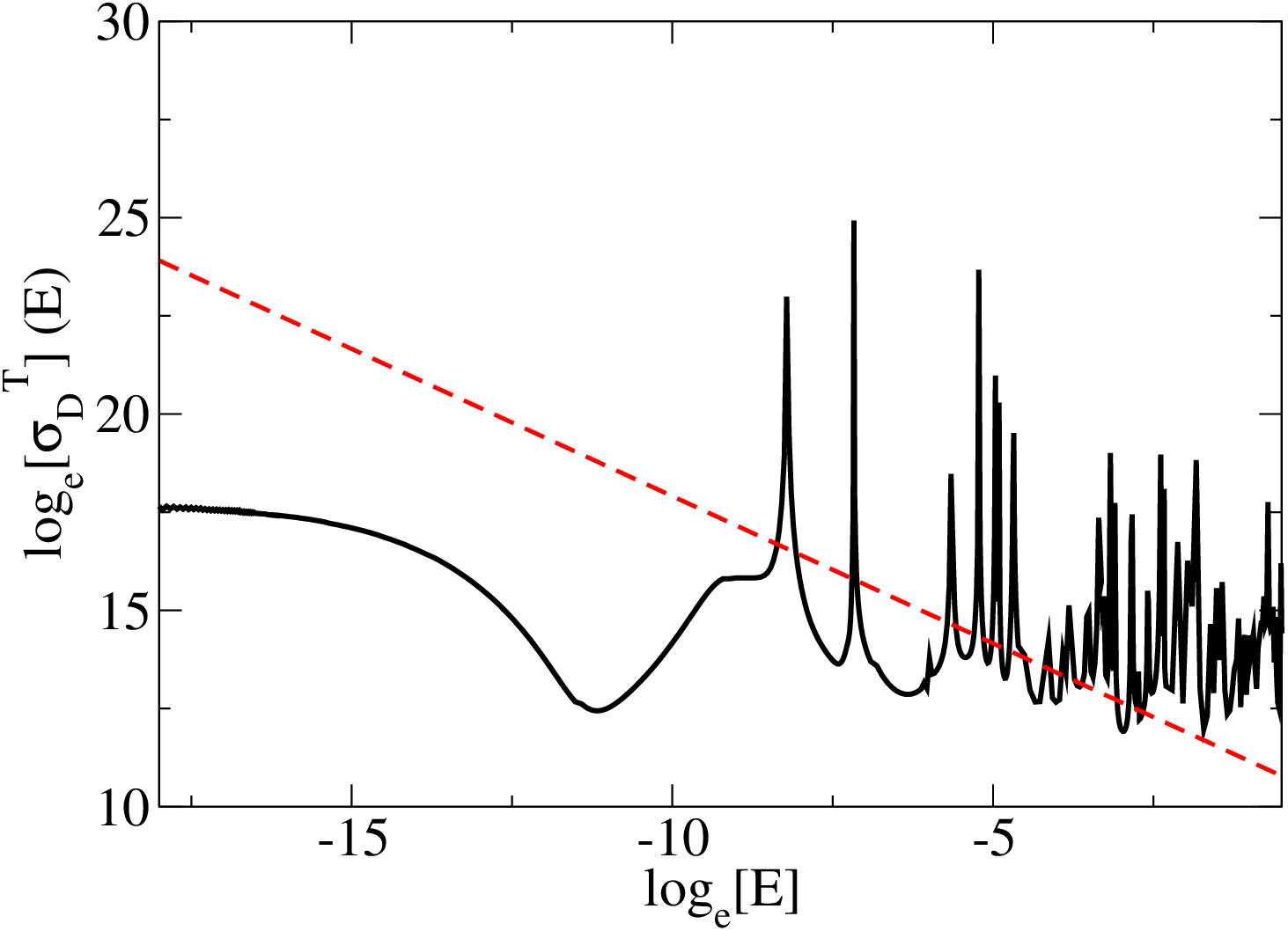}
         \label{fig:three sin xn}
     \end{subfigure}
     \caption{Diffusion cross sections are plotted as a function of energy for the potential X$^1$$\Sigma^+$ or a$^3$$\Sigma^+$, respectively. }
        \label{dc}
\end{figure}

\subsection{Diffusion cross section}
A charge or ion immersed in a dilute gas of atoms will diffuse through the stochastic scattering process. Therefore the diffusion of the ion in a gas of atoms will change the position of the ion with time, resulting  loss of forward momentum of the ion. This effect is quantified through diffusion cross-section or momentum transfer cross-section. The diffusion of the ion is characterised by diffusion coefficient. Diffusion occurs due to the binary collision of the ions with the atoms in the gas.  The total diffusion cross-section is associated with both elastic and inelastic contributions in the presence of atomic gases. It may be presumed that the inelastic diffusion cross-sections will be small compared to the corresponding elastic quantities 
By eliminating inelastic phenomena, the diffusion cross-section is associated with phase shifts for elastic collisions. The diffusion cross section for a hetero nuclear IA pair approaching along a single potential curve X$^1$$\Sigma^+$ or a$^3$$\Sigma^+$ is given by \cite{cote2016ultracold}
\begin{equation}
    \sigma_D^{S,T} (E) = \frac{4\pi}{k^2}\sum_{l=0}^{\infty}(2l+1)\sin^2\left(\eta_l^{S,T}-\eta_{l+1}^{S,T}\right)
\end{equation}

By using the above equation we evaluate diffusion cross sections ($\sigma_D^{S,T}$) for (BaLi)$^+$ system. In the left-hand and right-hand panels of FIG. \ref{dc}, we present ($\sigma_D^{S,T}$) as a function of energy in the log-log scale for the two concerned potentials. It is to be mentioned that for the homo nuclear system, the deviation of the Langevin behavior from the charge exchange cross sections (instead of spin exchange) can be expected when the scattering lengths are comparable in magnitudes and the same sign. To the best of our knowledge, there is no experimental data available in literature for the scattering lengths of  (BaLi)$^+$ system to justify the above arguments.

\section{Conclusion}
\label{sec-conclude}
We have studied the collisional properties of (BaLi)${^+}$ IA system which is of particular interest as the experimentally achievable cold Ba$^+$-Li combination corresponds to the ground asymptote and thereby protected against radiative charge transfer loss. We have made use of the \textit{ab initio} method to obtain the molecular potentials. We have presented a schematic description to study FR with an aim of venturing into the same in near future. We have calculated the elastic scattering cross-section at various regimes of collisional energy and verified its obedience to the semiclassical (-1/3) power law at high energies. We have studied the SE collision cross-section and compared its energy-dependency with the Langevin cross section which goes well in an expected intermediate energy regime. We have studied the PWPL effect for low partial waves and given an approximate formula for SE cross section in the PWPL scenario. Finally, we have investigated the diffusion properties of the concerned IA system.

\bibliography{sample.bib}

\end{document}